\documentclass[
    ,final            
  ]
  {aipproc}

\layoutstyle{6x9}

\begin{document}

\title{Setting the Triggering Thresholds on Swift}

\author{Kassandra M. McLean}
{address={Los Alamos National Laboratory}
  ,altaddress={University of Texas at Dallas}
}

\author{E. E. Fenimore}
{address={Los Alamos National Laboratory}}

\author{David Palmer}
{address={Los Alamos National Laboratory}}

\author{S. Barthelmy}
{address={Goddard Space Flight Center}}
\author{N. Gehrels}
{address={Goddard Space Flight Center}}
\author{H. Krimm}
{address={Goddard Space Flight Center}}
\author{C. Marwardt}
{address={Goddard Space Flight Center}}
\author{A. Parsons}
{address={Goddard Space Flight Center}}

\begin{abstract}The Burst Alert Telescope (BAT) on 
Swift has two main types of ``rate'' triggers: short and
long.  Short trigger time scales range from 4ms to 64ms, while
long triggers are 64ms to $\approx 16$ seconds.  While both short and long
trigger have criteria with one background sample (traditional
``one-sided'' triggers), the long triggers can also have criteria with
two background samples (``bracketed'' triggers) which remove trends in
the background.  Both long and short triggers can select
energy ranges of 15-25, 15-50, 25-100 and 50-350 KeV.  There are
more than 180 short triggering criteria and approximately 500 long
triggering criteria used to detect gamma ray bursts.  To fully
utilize these criteria, the thresholds must be set correctly.  The
optimum thresholds are determined by a tradeoff between avoiding false
triggers and capturing as many bursts as possible. We use realistic
simulated orbital variations, which are the prime cause of false
triggers.
\end{abstract}

\maketitle


\section{Introduction}

Swift is a rapidly slewing satellite, that can quickly point the field
of views (FOVs) of the X-Ray
telescope (XRT) and the ultraviolet-optical telescope (UVOT) at the
gamma-ray bursts (GRBs).  In order to begin this process, the Burst Alert
Telescope (BAT) must detect that a burst has occurred and locate its position.
Swift's behavior in responding to GRBs depends crucially on BAT
triggering, which means that it is vital to optimally set the thresholds for the
various trigger criteria.  The main drive behind the thresholds is one
of balancing sensitivity against false triggers.
This paper presents the procedure we used to select the thresholds for
Swift's various trigger criteria.

BAT uses about 800 different criteria to detect GRBs, each defined by a large number of commandable
parameters.  Usually the critical
parameter is the  time scale of the sample being analyzed for a statistically significant increase.  
There are three triggering systems. One is called
the ``short triggers'' and covers times ranging
from 4ms to 64 ms.  The short triggers have a fixed
background sample duration of about 1 sec.  The second system is the ``long triggers'' and covers 
time scales range from
64ms to as large as we dare command without trends in the background
producing too many false triggers($\approx 16$ sec currently).
There is a third trigger system (which we will not discuss) that searches for new point sources in images that 
are reconstructed periodically (typically every 32 sec, 320 sec, and \~1000 sec) from the detector plane 
observations. (See \cite{palmer_santafe} for details on the imaging.)

Both short and long triggers can target the 15-25, 15-50, 25-100 and
50-350 KeV energy ranges.  A criterion can also target a quadrant (or any combinations of quadrants)
of the detector plane, primarily to be more sensitive to bursts at the edge of the FOV. 
Swift has the most comprehensive set of
triggering criteria ever attempted.


We trigger on a burst when there is a statistically significant increase in the counts for any particular
criteria.  See \cite{fenimore_trigger} for more information on how the
trigger is evaluated.  Once the BAT has a rate trigger, it then images
the detector plane to find any new point sources.  If no new sources
are found, the trigger is rejected as false.  Thus, we can tolerate
false triggers in orbit and be more sensitive to bursts.

To create a high fidelity simulation of BAT in its orbital conditions
we simulate the steady, diffuse 
x-ray/gamma-ray sources, GRBs, and particle variations throughout
orbit. Two software
packages are used to accomplish this. The GRMCFLIGHT program simulates the gamma-ray transport through BAT.  GRMCFLIGHT
produces the photon energy, location, and time of incidence on the detector plane.  
GENERATE1355 is a program which simulates the BAT electronics, and produces the data
stream which is fed to the flight code. The combination of GRMCFLIGHT and GENERATE1355 allows us to trace
each photon from its origin (an astronomical source or the background) all the way through the BAT flight
software.

The ability to inject BATSE burst profiles into GRMCFLIGHT allows us to go even further and test the
high-level scientific features of the BAT flight software: namely triggering and imaging. This can then be
used for system-wide BAT tests, such as slewing to a GRB location. The result of these tests tells us how
many bursts we will trigger on, image, and slew to (see \cite{fenimore_ability}).

To set the thresholds, we run simulations with no injected bursts under various background conditions. We 
typically run between 10 and 20 hours of background, and try to set the threshold such that there would be 
no false 
triggers over this period of time. All simulations have a diffuse x-ray background component of about 8 KHz 
plus an additional 4 kHz background due to particles. We add various bumps in the background to mimic the 
orbital variations of the particle background. These bumps were typically Gaussian in shape with a full 
width at half maximum of about 24 minutes. We studied, in particular, bumps that added 20 kHz and 4 MHz. The 
4 MHz bump is what we expect in the SAA.


\section{Setting the Short Trigger Thresholds}

We ran each of the 180 short trigger criteria separately through the three different background variations: 
flat  12 KHz, flat 12 KHz plus a 20 KHz bump, and flat 12 KHz plus a 4 MHz bump.
Since
the background for the short triggers is within ~1 sec of the
foreground sample, we find that the thresholds are approximately the same for all three background variations. 
In order to
allow for statistical variations, we  ran each short trigger (i. e., 180 criteria) through about 10 orbits
(each with a 4 MHz SAA bump) to determine the  maximum score.
The short criteria tended to require about the same threshold for various permutations 
of areas of the detector plane and energy ranges. We set the thresholds for the short criteria to only 
depend on the time scale (i.e., 4, 8, 16, 32, and 64 msec). In Figure 1, the short criteria are denoted by 
filled squares and have thresholds of about $6 \sigma$ to $7 \sigma$.

\section{Setting the Long One-Sided Trigger Thresholds}

Swift uses two types of triggers in the > 64ms range: {\em
one-sided} and {\em bracketed} triggers.  One-sided triggers have only
one background sample which is before the foreground sample, and they
have been used in all past missions, such as BATSE. 
For the one-sided triggers, since they do not remove trends in the data, we
set them to withstand orbital variations of about 20kHz, a rather large bump for Swift in a 
quiet, non-SAA orbits.  

We have diagnostic reports to the telemetry of the 
maximum score for each criterion, each orbit. We plan to use this diagnostic output to set the thresholds on orbit 
once we have a few days of background data. To set the long thresholds on the ground prior to launch we simulate 14 
orbits (about 20 hrs) of background with 20kHz bumps in each
orbit. The telemetry provides a maximum score for each criteria in
each orbit. Let 
$S_{\rm max, i}$ be the overall maximum score for the $i$-th 
criteria seen over the 14 orbits. 
Let $\sigma_i$ be the RMS of the 14 maximum scores from each orbit.   We set $T_i$, the  threshold for the $i$-th 
criteria, to be
\begin{equation}
T_i = S_{\rm max, i}+2\sigma_i~~.
\label{THRESHOLD}
\end{equation}

The open squares in Fig.1 show the thresholds for the one-sided long triggers. The one-sided criteria have foreground 
durations of 0.128, 0.256, 0.512, and 1.024 sec. The one-sided criteria are very susceptible to trends in the 
background. A one percent trend over one second will produce a one sigma increase in the count rate. For that reason, 
we could not make one-sided criteria use foregrounds much longer than one second. 

For each foreground duration, there can be 36 different criteria. There are four different energy ranges (15-25, 15-50, 
25-100, 50-350 KeV) and nine combinations of sub-areas of the detector plane (the four quadrants, four halves, and the full 
detector plane). We do not necessarily use a set of criteria that
include all 36 permutations.

\section{Setting Long Bracketed Trigger Thresholds}

Bracketed triggers have a background sample before and after the foreground sample,
allowing a fit to trends in the background rate.  To date these have only been
used in the HETE flight trigger algorithm.
The bracketed triggers efficiently remove background trends so we
could set the thresholds to be much lower than the one-sided triggers
and use much longer foreground time samples.
They are occasionally sensitive to the {\em peak} of a bump in the
background.  We  initially simulated bumps ranging
from 200Hz to 2 MHz in steps of a factor of ten.

The opened circles in Fig. 1 give the thresholds for various permutations of energy and detector plane regions for 
foreground durations between 0.128 and 16 sec.  We used 14 orbits of background with a flat background of 20 KHz. No 
bumps were used because the two-sided criteria removes trends and the size of the bump had little effect on the 
thresholds.   Equation \ref{THRESHOLD} was applied to the maximum scores reported by 
the diagnostic software to obtain the thresholds.
By using a flat background, we were able to get scores in the range of
$3 \sigma$ to $5 \sigma$.

\section{Conclusion}

\begin{figure}
  \includegraphics[height=.3\textheight]{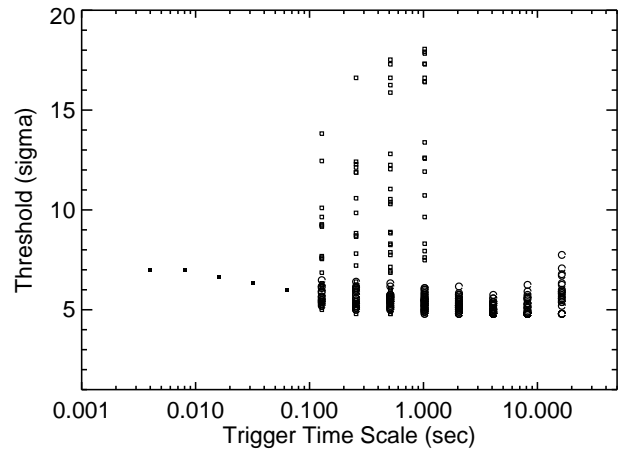}
  \caption{Thresholds for the BAT rate trigger criteria as a function of the foreground duration. The solid squares are 
for the short rate triggers.  The open squares are for one-sided long criteria and the open circles are for two-sided 
long criteria. In contrast, the equivalent threshold for BATSE was 11
sigma and foreground time durations were only 64, 256, and 1024 ms. 
}
\end{figure}

Our strategy is to have the trigger thresholds as low as possible and allow approximately 2 to 3 false triggers per 
hour. The flight software will form an image at the time of the trigger and those false triggers will be rejected 
because there will not be a new point source in the image. Sources in a coded aperture image will usually have a 
smaller signal-to-noise than the signal-to-noise in the trigger. As a result, valid triggers near the threshold will 
not have significant new point sources in the image.  Thus although
Fig. 1 shows the thresholds required for a trigger, a successful image
requires a stronger signal.   

There is often the concern that the two-sided criteria introduce a delay in recognizing that a burst is occurring. 
This is true. Our strategy is that we will use one-sided criteria with
durations as long as we can (which is $\approx 1$ 
sec) and with threshold that avoid excessive false triggers. Those
thresholds, unfortunately, can be as large as $\approx 15$. 
The two-sided criteria indeed take longer, but they are only used when the one-sided were unable to detect the 
burst. It is better to get the burst later than not at all.







\end{document}